\documentclass[smallabstract,smallcaptions]{dccpaper}
\usepackage{geometry}

\makeatletter
\renewcommand{\@biblabel}[1]{[#1]}
\makeatother
\makeatletter

\def\Section#1{\vspace*{-0.125in}\section{#1}}
\def\SubSection#1{\vspace*{-0.125in}\subsection{#1}}

\usepackage{suffix}
\WithSuffix\def\Section*#1{\vspace*{-0.125in}\section*{#1}}
\WithSuffix\def\SubSection*#1{\vspace*{-0.125in}\subsection*{#1}}

\makeatother

\usepackage{epsfig}
\usepackage{pgfplots}

\definecolor{nodeblue}{RGB}{209, 236, 250}

\graphicspath{ {./figures/} }
\DeclareGraphicsExtensions{.pdf,.png}

\newlength{\figurewidth}
\newlength{\smallfigurewidth}

\usepackage{amsmath}
\usepackage{amssymb}

\usepackage{booktabs}
\usepackage{tabularx}
\usepackage{multirow}
\usepackage[font=small]{caption}

\newcolumntype{L}[1]{>{\raggedright\let\newline\\\arraybackslash\hspace{0pt}}p{#1}}
\newcolumntype{C}[1]{>{\centering\let\newline\\\arraybackslash\hspace{0pt}}p{#1}}
\newcolumntype{R}[1]{>{\raggedleft\let\newline\\\arraybackslash\hspace{0pt}}p{#1}}

\usepackage{csquotes}
\usepackage[vietnamese,UKenglish]{babel}

\usepackage{color}
\usepackage[hidelinks]{hyperref}
\usepackage{url}
\usepackage[binary-units=true]{siunitx}
\usepackage{cleveref}
\usepackage{todonotes}

\usepackage{abbreviations}
\begin{document}

\title
{\large
\textbf{Making compression algorithms for Unicode text}
}

\author{%
Adam Gleave \\
\emph{University of Cambridge}\\
\href{mailto:adam.gleave@cantab.net}{\nolinkurl{adam.gleave@cantab.net}}
\and
Christian Steinruecken\\
\emph{University of Cambridge}\\
\href{mailto:tcs27@cam.ac.uk}{\nolinkurl{tcs27@cam.ac.uk}}
\vspace*{1em}
}

\maketitle
\thispagestyle{empty}

\begin{abstract}
The majority of online content is written in languages other than English, and is most commonly encoded in \utfeight, the world's dominant Unicode character encoding.
Traditional compression algorithms typically operate on individual bytes. While this approach works well for the single-byte \ascii encoding, it works poorly for \utfeight, where characters often span multiple bytes.
Previous research has focused on developing Unicode compressors from scratch, which often failed to outperform established algorithms such as \bziptwo.

We develop a technique to modify byte-based compressors to operate directly on Unicode characters, and implement variants of \lzw and \ppm that apply this technique.
We find that our method substantially improves compression effectiveness on a \utfeight corpus, with our \ppm variant outperforming the state-of-the-art \ppmii compressor.
On \ascii and binary files, our variants perform similarly to the original unmodified compressors.
\end{abstract}

\Section{Introduction} \label{sec:intro}
The Unicode encoding scheme \utfeight is the most commonly used character set on the Internet, adopted by 87.0\% of websites~\cite{w3techs:encoding}.
Whereas the legacy \ascii encoding represents each character by a single byte, \utfeight maps non-\ascii characters to sequences of two to four bytes.
The majority of websites are now in languages other than English~\cite[table~8]{pimienta:2009}, and contain multi-byte \utfeight characters.
Text compression has yet to adapt to this shift, as most compressors still operate on individual bytes.
\enlargethispage{1em}

To address this deficit, The Unicode Consortium defined the Standard Compression Scheme for Unicode (\scsu) and Binary Ordered Compression for Unicode (\bocu)~\cite{wolf:2005,scherer:2006}.
Both methods exploit the property that adjacent characters in a text tend to have nearby code points.\footnote{The Unicode code space is divided into \emph{blocks} of characters that belong to the same script, and most texts use characters from only a small number of blocks.}
{\hyphenation{compactly}Although these Unicode-aware compressors are useful for compactly representing short strings, Atkin and Stansifer~\cite{atkin:2003} find that on longer texts they are outperformed by \gzip and \bziptwo.}
Using \scsu as a preprocessing step improves \gzip's results slightly,
but unmodified \bziptwo is still the winner.

In this paper, we investigate an alternative approach. Like Atkin and Stansifer, we make use of existing compression algorithms: in our case, \ppm and \lzw. However, instead of preprocessing the input, we modify these algorithms to operate directly over Unicode characters rather than bytes.
This approach is challenging: there are far more Unicode characters than bytes, making it harder to learn the source distribution.
We propose an adaptive character model based on \emph{P\'{o}lya trees} that is well-suited for learning Unicode character distributions.

Our approach yields a 12.2\% average improvement in compression effectiveness for \lzw over a corpus of \utfeight files. 
{\hyphenation{improvement}This result compares favourably with the Unicode-aware \lzseventyseven variant from Fenwick and Brierley that achieved only a 2\% improvement~\cite{fenwick:1998}. Unfortunately their corpus is not available to us, precluding a direct comparison.}

For \ppm, our approach gives a 6.1\% improvement on our \utfeight corpus, and outperforms Shkarin's best-in-class \ppmii implementation~\cite{shkarin:2006a}. Moreover, compression effectiveness of our \lzw and \ppm variants over the Canterbury corpus~\cite{arnold:1997} is no worse than for the byte-based versions of \lzw and \ppm.

\Section{A token representation for \SAFEutfeight}
\null\vspace{-1ex}%
\SubSection{Background on \SAFEutfeight} \label{sec:utf8}

The Unicode Standard~\cite{unicode:2016} defines how to encode multilingual text.
The latest version at the time of writing, Unicode 9.0.0, contains \num{128172} characters from all the world's major writing systems, including contemporary and historic scripts.
Each character is identified by a unique integer, called the character's \emph{code point}.

Unicode defines three \emph{encoding schemes} that map code points to byte sequences: \utfeight, \utfsixteen and \utfthirtytwo.
For the interchange of Unicode text, \utfeight is the \emph{de facto} standard.
{\hyphenation{programming}Fewer than 0.1\% of websites are encoded in \utfsixteen or \utfthirtytwo~\cite{w3techs:encoding}, although these encoding schemes are often used as an internal representation in programming language APIs.}

\begin{table}
\centering
\begin{tabular}{rllll}
\toprule
\textbf{Code point} & \textbf{Byte 1} & \textbf{Byte 2} & \textbf{Byte 3} & \textbf{Byte 4} \\
\midrule 
\code{0xxxxxxx} & \code{0xxxxxxx} & & & \\
\code{yyy yyxxxxxx} & \code{110yyyyy} & \code{10xxxxxx} & & \\
\code{zzzzyyyy yyxxxxxx} & \code{1110zzzz} & \code{10yyyyyy} & \code{10xxxxxx} & \\
\code{uuuuu zzzzyyyy yyxxxxxx} & \code{11110uuu} & \code{10uuzzzz} & \code{10yyyyyy} & \code{10xxxxxx} \\
\bottomrule 
\end{tabular}
\caption[\SAFEutfeight byte sequences]{The mapping between code points and byte sequences for \utfeight~\cite[table~3.6]{unicode:2016}.}
\vspace*{-1em}
\label{table:utf8:mapping}
\end{table}

\utfeight is a variable-length encoding scheme, mapping code points to \emph{code words} of one to four bytes, as specified in \cref{table:utf8:mapping}. Code points between \unicodechar{00} and \unicodechar{7F} are mapped to the \ascii character of the same value. This property is called ``\ascii transparency'', and is a key reason for the popularity of \utfeight. Another desirable property is that \utfeight is a self-synchronising code, since the range of valid values for the first byte in a sequence is disjoint from those of the trailing bytes.

Any byte sequence that does not match one of the rows in \cref{table:utf8:mapping} is ill-formed.
Some other ill-formed sequences also exist (e.g.~code points above \unicodechar{10FFFF}
and code points from \unicodechar{D800} to \unicodechar{DFFF}),
as described in the RFC specification of \utfeight~\cite{RFC3629}.

\SubSection{A token representation} \label{sec:utf8:token}
We developed an invertible \utfeight decoder and encoder that maps between sequences of bytes and sequences of \emph{tokens}.
There are three types of tokens:
  \tchar{$c$} represents a Unicode character with code point $c$,
  \tbyte{$b$} represents a byte~$b$ in an ill-formed sequence, and
  \teof\ is an end-of-file marker.

\utfeight text is transformed to a sequence of \tchar{$c$} tokens, terminated by~\teof.
If decoding fails on byte~$b$ then \tbyte{$b$} is emitted.\footnotemark{} Decoding continues without interruption on the next byte, because \utfeight is a self-synchronising code.
The transform is reversible: any sequence of bytes (not just valid \utfeight) can be transformed into a sequence of tokens and back.
This paper describes several techniques for compressing the resulting token sequences.
\footnotetext{It is possible to extend this scheme to specify the exact cause of a decoding error~\cite[section~2.3]{gleave:2016}. However, our tests find little difference in the resulting compression effectiveness, so we favour the simpler approach.}

For computational convenience, our implementation represents tokens as integers. Characters are represented by their code point, with bytes following sequentially afterwards, and~\teof at the end.

\Section{Contextless models over tokens} \label{sec:symbol-models}

A model may be \emph{adaptive} (learning from past input symbols),
or \emph{non-adaptive} (a~fixed probability distribution).
This section introduces three contextless models over \emph{tokens} as defined in \cref{sec:utf8:token}.
Two of these models are non-adaptive and one is adaptive.
All three models are used as components in our Unicode variants of \lzw and \ppm (described in section~\ref{sec:complex-models}),
where they serve as base models for unseen symbols.

\SubSection{Non-adaptive: the uniform and \SAFEutfeight distributions} \label{sec:symbol-models:static}
The uniform distribution is a simple example of a non-adaptive model,
assigning equal probability mass to each symbol in the alphabet.
However, a uniform distribution over the Unicode alphabet is a poor choice if we believe that some scripts (such as the Latin alphabet) are consistently more likely than others (such as Egyptian hieroglyphs).

When compressing \utfeight input, a natural base model to use is the probability distribution implicitly defined by the \utfeight encoding itself. 
This distribution assigns each Unicode character $c$ the probability mass $\p{c} = k/256^l$, where $l$ is the length (in bytes) of the code word that $c$ is mapped to and $k$ is a normalising constant (needed because \utfeight is not a compact representation). 
Characters with lower code points tend to occur more frequently, so are allocated to shorter code words by \utfeight.

A non-adaptive model can never be a good fit for all texts. 
In particular, texts are typically written in a single script, resulting in the probability mass being \emph{concentrated} in the contiguous region of code points assigned to that script.
In the next section, we describe an adaptive character model
that exploits this property when learning the source distribution.

\SubSection{Adaptive: the P\'{o}lya tree learner} \label{sec:symbol-models:polya}
\begin{figure}
\centering
\begin{tikzpicture}[
       node/.style={draw, fill=nodeblue, text=black, text centered, minimum width=1cm},
       level 1/.style={circle, sibling distance=40mm},
       level 2/.style={rectangle, sibling distance=20mm, minimum width=1cm, minimum height=0.5cm, label distance=0.2cm}
]
\node [node, circle] (root) { $4 : 2$ }
	child {node [node] (l) { $3 : 1$ }
		child {node [node,label=below:{$\frac{20}{48}$}] (a) { A } edge from parent node[above left] {$\frac{2}{3}$}}
		child {node [node,label=below:{$\frac{10}{48}$}] (b) { B } edge from parent node[above right] {$\frac{1}{3}$}}
		edge from parent node[above left] {$\frac{5}{8}$}
	}
	child {node [node] (r) { $1 : 1$ }
		child {node [node,label=below:{$\frac{9}{48}$}] (c) { C } edge from parent node[above left] {$\frac{1}{2}$}}
		child {node [node,label=below:{$\frac{9}{48}$}] (d) { D } edge from parent node[above right] {$\frac{1}{2}$}}
		edge from parent node[above right] {$\frac{3}{8}$}
	};
\end{tikzpicture}
\caption[A finite P\'{o}lya tree]{A finite P\'{o}lya tree over the alphabet $\alphabet = \{A,B,C,D\}$ after 6 symbol observations.
The parameters of the Beta prior at each internal node~$i$ were set to $\alpha_i = \beta_i = 1$.
Internal nodes are labelled with $L_i : R_i$, where $L_i$ is the number of times the left branch was taken, and $R_i$~the number of times the right branch was taken.
Edges are labelled with the posterior branching probability.
Each leaf node $x \in \alphabet$ is labelled with its predictive probability $\p{x_7\?=x \| \vc{x}{1}{6}, \boldsymbol{\alpha},\boldsymbol{\beta}}$ from~\eqref{eq:polya-adapt-final}.}
\label{fig:single-symbol:polya}
\end{figure}
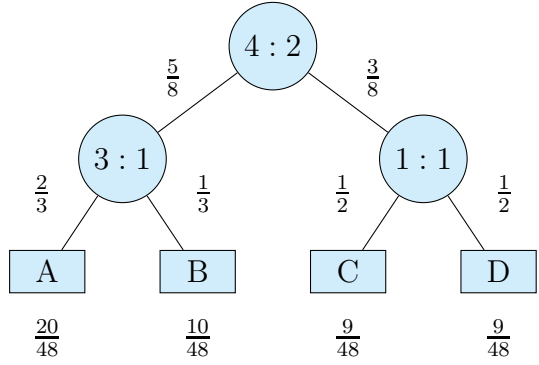

Consider a balanced binary search tree whose leaf nodes contain the symbols of the alphabet~$\alphabet$.
Suppose each internal node $i$ has an associated probability $\theta_i$
of choosing the left branch over the right branch.
Each symbol $x \in \alphabet$ can be uniquely identified by the path of branching decisions from the root of the tree. The probability of symbol~$x$ is defined as the joint probability of these branching decisions:
\def\pathprod#1#2#3{\kern#3\prod_{#1 \in \ppath{#2}}\kern#3}%
\begin{equation}\label{eq:polya-non-adapt}
  \p{x \| \boldsymbol{\theta}}
     \ =\  \pathprod{i}{x}{-1.5ex} \Bern{b_i \| \theta_i}
     \ =\  \pathprod{i}{x}{-1.5ex} {\theta_i}^{b_i} \left(1-\theta_i\right)^{1-b_i}.
\end{equation}
Here $\ppath{x}$ is the set of internal nodes from the root to symbol $x$,
$b_i \in \{0,1\}$ is the branching decision taken at node $i$,
and $\boldsymbol{\theta}$ is the collection of all $\theta_i$.

The node branching biases $\boldsymbol{\theta}$ can be learned adaptively.
To derive such a method, we place conjugate priors $\theta_i \sim \Beta{\alpha_i, \beta_i}$ on
each $\theta_i$, and maintain two counts $L_i$ and $R_i$ of how often the left and right branches were taken at node $i$.
The posterior over $\theta_i$ (given $\alpha_i$, $\beta_i$, $L_i$, and $R_i$)
is a Beta distribution with parameters $\alpha_i\+L_i$ and $\beta_i\+R_i$.
The $\theta_i$ can then be integrated out,
resulting in the following closed form expression for the posterior symbol probability:
\begin{equation} \label{eq:polya-adapt-final}
  \p{x_{k+1} \| \vc{x}{1}{k}, \boldsymbol{\alpha}, \boldsymbol{\beta}}
  \ =\  \pathprod{i}{x_{k+1}}{-2.3ex}
           \Bern{ b_i \| \frac{\alpha_i + L_i}{\alpha_i + \beta_i + L_i + R_i} },
\end{equation}
where $\vc{x}{1}{k}$ is the collection of the first $k$ symbol observations,
and $\boldsymbol{\alpha}$ and $\boldsymbol{\beta}$ are the collections of all $\alpha_i$ and $\beta_i$. The initial predictive distribution before any symbols are observed, $\p{x_1 \| \boldsymbol{\alpha}, \boldsymbol{\beta}}$, can be made equal to any chosen distribution by an appropriate choice of $\boldsymbol{\alpha}$ and $\boldsymbol{\beta}$.
Learning a new symbol $x'$ simply requires incrementing the matching $L_i$ or $R_i$ counters at each node $i \in \ppath{x'}$.
An example P\'{o}lya tree is depicted in~\cref{fig:single-symbol:polya}.

P\'olya trees have the interesting property that learning
a symbol $x$ not only boosts the predictive probability of $x$
(as conventional histogram learning methods do),
but also other symbols that are close to $x$ in the tree.
This property is apparent in \cref{fig:single-symbol:polya}: \sym{B}~is assigned a higher probability than~\sym{C} and~\sym{D} (despite all three symbols occurring once each), because~\sym{B} is next to the frequently occurring symbol~\sym{A}.

A P\'{o}lya tree can learn any symbol distribution given enough observations, but they are especially fast at learning distributions where symbols in contiguous regions
have similar occurrence probabilities.
Such distributions often occur in Unicode texts.
The P\'olya tree character model was first introduced in~\cite[section~4.3.1]{steinruecken:2014}.

\Section{Compressing \SAFEutfeight text} \label{sec:complex-models}
We extend two widely used compressors -- \ppm and \lzw\ -- to operate over tokens instead of bytes, and escape to a base model the first time each symbol is observed.

\SubSection{\SAFEppm} \label{sec:complex-models:ppm}
The Prediction by Partial Matching (\ppm) algorithm~\cite{cleary:1984} encodes symbols using context-dependent symbol distributions via arithmetic coding~\cite{witten:1987}.
It is parametrised by a maximum context depth~$d$. To encode a symbol~$x$, \ppm starts with the $d$-length context preceding~$x$.
(Unless $x$ is one of the first $d$ symbols, in which case it uses as many preceding symbols as are available.) 
If~$x$ is unseen in this context, an \emph{escape symbol} \code{ESC} is encoded (a virtual symbol that is
treated as part of the alphabet).
The context is then shortened to length $d-1$ by dropping the symbol furthest away from~$x$, and the process is repeated recursively.

In most cases, a context is eventually reached where $x$~has been seen before, and $x$~is then encoded with the histogram learned for this context.
However, when~$x$ occurs in the input data for the first time, $x$~will be absent even in the empty context of length~$0$.
In this case, after encoding a final \code{ESC}, $x$ is transmitted using a uniform distribution.
Our implementation generalises this construction by allowing base models other than the uniform distribution.
In particular, we recommend using the P\'{o}lya tree as a base model, which can reduce the cost of unseen symbols.

{\hyphenation{context}
This generic description of \ppm does not specify how the histogram learned in each context is used to assign probabilities to symbols.
Cleary and Witten~\cite{cleary:1984} state that ``in the absence of \emph{a priori} knowledge, there seems to be no theoretical basis for choosing one solution [method] over another''.
Different methods have been proposed, such as \ppma and \ppmb~\cite{cleary:1984}, \ppmc~\cite{moffat:1990} and \ppmd~\cite{howard:1993}.}
These methods were designed for \ppm compressors operating over byte alphabets, and they might not perform as well over the much larger space of tokens.

We decided to base our experiments on \ppmg, a method with two continuous parameters that generalises \ppma and \ppmd~\cite[section~6.4.2]{steinruecken:2014}.
\ppmg has a discount parameter ${\beta \in [0,1]}$ and a concentration parameter ${\alpha \in [-\beta,\infty)}$, and assigns the following symbol and escape probabilities:
\begin{equation}
\p{x} = \frac{n_x - \beta}{N + \alpha}\cdot \mathbb{I}\left[n_x > 0\right] \qquad \text{and} \qquad \p{\text{\code{ESC}}} = \frac{U\beta + \alpha}{N + \alpha}.
\end{equation}
Here $n_x$~is the number of observations of~$x$ in the current context, $N$~is the total number of observations in the current context and~$U$ is the number of unique symbols observed in the current context.%
\footnote{%
  During learning, the symbol counts $n_x$ are updated using the standard exclusion method,
  as pioneered by Moffat~\cite{moffat:1990} in his ``second modification'' to \ppm.
}

The $\alpha$ and $\beta$ parameters have a similar interpretation to those in a Pitman--Yor process~\cite[section~6.6.3]{steinruecken:2014}.
Informally, the parameter~$\alpha$ can be viewed as a pseudo-count for the escape symbol \code{ESC}.
The discount parameter $\beta$ influences the tail behaviour of the symbol distribution,
allowing the escape probability to vary based on~$U$, the number of unique symbols observed in the context.

Although we agree with Cleary and Witten that there is no way to design a method without making some \emph{a priori} assumptions, the optimal parameters for \ppmg can be determined \emph{a posteriori} from training data.
These optimal parameter settings may be useful for compressing files that are similar to the training corpus.
We expected \ppmg's optimal parameters for byte-based and token-based alphabets to be very different.
Surprisingly, we found this not to be the case, as can be seen in~\cref{table:eval:optimal-parameters}.

\SubSection{\SAFElzw} \label{sec:complex-models:lzw}
The Lempel--Ziv--Welch algorithm (\lzw) is a dictionary compressor~\cite{welch:1984}.
While \ppm typically compresses more effectively than \lzw, especially on text, \lzw and other dictionary compressors are often more computationally efficient.
\lzw makes for an interesting case study as a representative of the wider class of dictionary coding algorithms.

Unconcerned by memory usage or execution speed, our experimental imple\-men\-tation of \lzw places no upper limit on the dictionary size $N$, and encodes each index in $\log_2 N$ bits using arithmetic coding~\cite{witten:1987}.
Traditional implementations initialise the dictionary to contain every byte.
This practice could be extended na\"ively to tokens, but at the cost of considerably inflating the size of the dictionary, and thus increasing the number of bits needed to encode each index.
This approach would be wasteful, as most tokens will never occur in most files.

We resolve this problem by escaping to a base model over tokens the first time a symbol is seen. The dictionary is initialised to contain a single entry~$\emptystring$, the empty string. When the compressor encounters an unseen symbol~$x$, it first encodes~$\emptystring$, and then uses the base model to encode~$x$. As with any substring seen for the first time, $x$~is added to the dictionary, so the escape procedure is used at most once for each symbol.
An approach similar to escaping was envisaged by Welch~\cite[page~11]{welch:1984}:
\begin{displayquote}\small\sl%
``The compression string table could be initialised to have only the null string. In that case, a special code is necessary for the first use of each single-character string. This approach improves compression if relatively few different symbols occur in each message.''
\end{displayquote}
However, he does not elaborate any further on this method. Our tests in \cref{sec:eval} find that escaping improves compression effectiveness even when using a byte alphabet.

\Section{Evaluation} \label{sec:eval}

\SubSection{Corpora}\label{sec:corpora}
\begin{table}
\begin{tabularx}{\textwidth}{lR{4em}l}
\toprule
\textbf{File} & \centering\textbf{Size (bytes)} & \textbf{Description} \\
\midrule
\code{ara-tabula.txt} & \num{2330891} & \emph{Tabula Rogeriana}, Arabic geographical text \\
\code{ben-kobita.txt} & \num{430218} & \emph{Shesher Kobita}, seminal Bengali novel \\
\code{hin-baital.txt} & \num{214533} & \emph{Baital Pachisi}, Indian legend in Hindi \\
\code{jav-tuban.txt} & \num{141694} & \emph{Rangsang Tuban}, early Javanese novel \\
\code{jpn-yujo.txt} & \num{203373} & \emph{Yujo}, Japanese novel \\
\code{lah-wiki.txt} & \num{124 083} & \emph{Wikipedia} article, in Punjabi \\
\code{por-noites.txt} & \num{115877} & \emph{Noites de Insomnia}, Portuguese book \\
\code{rus-mosco.txt} & \num{82039} & \emph{Moscovia v Predstavlenii Inostrantsev}, Russian work \\
\code{spa-trans.txt} & \num{311333} & \emph{Transfusi\'{o}n}, Spanish novel \\
\code{zho-you.txt} & \num{64526} & \emph{You Xue Qiong Lin}, classical Chinese text \\ %
\code{mix-sake.txt} & \num{257061} & Japanese article, with parallel English translation \\
\bottomrule
\end{tabularx}
\caption[Our corpus]{%
  Our \utfeight corpus, containing a sample of the world's most common languages.
}%
\label{table:eval:our-corpus}
\end{table}

\begin{table}
\centering%
\begin{tabular}{ccccrrc}
\toprule
\textbf{Label} & \textbf{Base} & \textbf{Alphabet} & \textbf{$\mathbf{d_{\text{opt}}}$} & \textbf{$\boldsymbol{\alpha_\text{opt}}$} & \textbf{$\boldsymbol{\beta_\text{opt}}$} \\
\midrule
\ppmuniformbyte  & Uniform   & Byte  &  6  &  0.095  & 0.409 \\
\ppmuniformtoken & Uniform   & Token &  5  & -0.001  & 0.514 \\
\ppmpolyatoken   & P\'{o}lya & Token &  5  &  0.000  & 0.513 \\
\bottomrule
\end{tabular}
\caption[Optimal \SAFEppmg parameter settings]{%
{\hyphenation{optimisation}Optimal settings of the \ppmg parameters $(d,\alpha,\beta)$ on training data.
These values are curiously close to $\alpha\?=0$ and $\beta\?=\frac12$,
the parameter settings that correspond to \ppmd~\cite{howard:1993}.
The training data and optimisation procedure are described by Gleave~\cite[section~6.4]{gleave:2016}.}
}
\label{table:eval:optimal-parameters} 
\end{table}

We assembled the corpus of \utfeight texts listed in \cref{table:eval:our-corpus}. First, we collected public-domain texts for all 22~languages with at least 50 million native speakers, excluding English, according to Ethnologue~\cite{ethnologue:2016}.
Next, we sampled ten languages (without replacement) with probability proportional to their number of speakers. For each sampled language we then selected the text with the lowest MD5 hash. We also included a multilingual text, \code{mix-sake.txt}. Our compressors were tested both on this corpus, and on the English-language Canterbury corpus~\cite{arnold:1997}.

\SubSection{Parameter selection}
We set the \ppmg parameters to values that maximised the compression effectiveness on training data.
These values are shown in \cref{table:eval:optimal-parameters}.
The optimal depth is $6$ for bytes, but only $5$ for tokens, since a token can substitute for multiple bytes.
The optimal~$\beta$ is higher under the token alphabet, because the number of unique tokens varies more among files than the number of unique bytes, which is always bounded by $256$. These differences highlight the importance of a principled method of parameter selection.

Our experiments found these parameters were almost the same as those that would be optimal on the test corpus described in section \ref{sec:corpora}.
Furthermore, a sensitivity analysis showed that \ppm is fairly robust to parameter choice.
The training corpus, our optimisation procedure, and details of these additional experiments are described by Gleave~\cite[section~6.4]{gleave:2016}.

The P\'{o}lya tree's $\boldsymbol{\alpha}$ and $\boldsymbol{\beta}$ parameters were set such that the initial predictive distribution is equal to the \utfeight implicit probability distribution.

\SubSection{Experimental results}
\begin{table}[p]
\setlength{\tabcolsep}{1ex}
\hspace{2em}
\begin{tabular}{m{0em}lrccccccl}
\toprule
& \textbf{File} & \multicolumn{1}{c}{\textbf{Size}} & \multicolumn{3}{c}{\SAFElzw} & \multicolumn{4}{c}{\SAFEppm}\\
 &  & \multicolumn{1}{c}{(KiB)} & \SAFElzwuniformbyte & \SAFElzwuniformstoken & \SAFElzwpolyastokenuniformbyte & \SAFEppmtraininguniformbyte & \SAFEppmtraininguniformstoken & \SAFEppmtrainingpolyastokenuniformbyte & \hspace{-5pt}\SAFEppmii\\
\midrule
\multirow{8}{*}{\hspace{-2em}\rotatebox{65}{\emph{\ascii}~}}
 & \code{alice29.txt} & \num{149} & {\kern-0.35em\colorbox[rgb]{0.42823530260254355,0.7729642580537235,0.6636217040174148}{\textcolor[rgb]{0,0,0}{3.154}}\kern-0.35em} & {\kern-0.35em\colorbox[rgb]{0.42823530260254355,0.7729642580537235,0.6636217040174148}{\textcolor[rgb]{0,0,0}{3.160}}\kern-0.35em} & {\kern-0.35em\colorbox[rgb]{0.42823530260254355,0.7729642580537235,0.6636217040174148}{\textcolor[rgb]{0,0,0}{3.153}}\kern-0.35em} & {\kern-0.35em\colorbox[rgb]{0.7419607996940613,0.9027297251364764,0.8689581008518443}{\textcolor[rgb]{0,0,0}{2.203}}\kern-0.35em} & {\kern-0.35em\colorbox[rgb]{0.7419607996940613,0.9027297251364764,0.8689581008518443}{\textcolor[rgb]{0,0,0}{2.188}}\kern-0.35em} & {\kern-0.35em\colorbox[rgb]{0.7482353091239929,0.9051903170697829,0.8725259591551388}{\textcolor[rgb]{0,0,0}{2.181}}\kern-0.35em} & {\kern-0.35em\colorbox[rgb]{0.7733333468437195,0.915032684803009,0.8867973923683167}{\textcolor[rgb]{0,0,0}{\kern-0.17em\textbf{2.101}\kern-0.17em}}\kern-0.35em}\\
 & \code{asyoulik.txt} & \num{122} & {\kern-0.35em\colorbox[rgb]{0.3658592928858364,0.7423298884840572,0.6006920583107892}{\textcolor[rgb]{0,0,0}{3.392}}\kern-0.35em} & {\kern-0.35em\colorbox[rgb]{0.3658592928858364,0.7423298884840572,0.6006920583107892}{\textcolor[rgb]{0,0,0}{3.398}}\kern-0.35em} & {\kern-0.35em\colorbox[rgb]{0.3658592928858364,0.7423298884840572,0.6006920583107892}{\textcolor[rgb]{0,0,0}{3.391}}\kern-0.35em} & {\kern-0.35em\colorbox[rgb]{0.6415686488151551,0.863360254203572,0.811872367999133}{\textcolor[rgb]{0,0,0}{2.502}}\kern-0.35em} & {\kern-0.35em\colorbox[rgb]{0.6541176676750183,0.8682814380701851,0.8190080846057218}{\textcolor[rgb]{0,0,0}{2.468}}\kern-0.35em} & {\kern-0.35em\colorbox[rgb]{0.6541176676750183,0.8682814380701851,0.8190080846057218}{\textcolor[rgb]{0,0,0}{2.461}}\kern-0.35em} & {\kern-0.35em\colorbox[rgb]{0.6980392336845398,0.8855055816033307,0.8439830927287831}{\textcolor[rgb]{0,0,0}{\kern-0.17em\textbf{2.340}\kern-0.17em}}\kern-0.35em}\\
 & \code{cp.html} & \num{24} & {\kern-0.35em\colorbox[rgb]{0.338546722426134,0.7275663368842181,0.5667358840213103}{\textcolor[rgb]{0,0,0}{3.512}}\kern-0.35em} & {\kern-0.35em\colorbox[rgb]{0.3294425322728999,0.722645153017605,0.5554171592581506}{\textcolor[rgb]{0,0,0}{3.556}}\kern-0.35em} & {\kern-0.35em\colorbox[rgb]{0.338546722426134,0.7275663368842181,0.5667358840213103}{\textcolor[rgb]{0,0,0}{3.506}}\kern-0.35em} & {\kern-0.35em\colorbox[rgb]{0.7043137431144715,0.8879661735366372,0.8475509510320776}{\textcolor[rgb]{0,0,0}{2.312}}\kern-0.35em} & {\kern-0.35em\colorbox[rgb]{0.6980392336845398,0.8855055816033307,0.8439830927287831}{\textcolor[rgb]{0,0,0}{2.335}}\kern-0.35em} & {\kern-0.35em\colorbox[rgb]{0.710588252544403,0.8904267654699438,0.851118809335372}{\textcolor[rgb]{0,0,0}{2.285}}\kern-0.35em} & {\kern-0.35em\colorbox[rgb]{0.7482353091239929,0.9051903170697829,0.8725259591551388}{\textcolor[rgb]{0,0,0}{\kern-0.17em\textbf{2.174}\kern-0.17em}}\kern-0.35em}\\
 & \code{fields.c} & \num{11} & {\kern-0.35em\colorbox[rgb]{0.3658592928858364,0.7423298884840572,0.6006920583107892}{\textcolor[rgb]{0,0,0}{3.394}}\kern-0.35em} & {\kern-0.35em\colorbox[rgb]{0.34309881750275106,0.7300269288175246,0.5723952464028901}{\textcolor[rgb]{0,0,0}{3.494}}\kern-0.35em} & {\kern-0.35em\colorbox[rgb]{0.3704113879624535,0.7447904804173637,0.606351420692369}{\textcolor[rgb]{0,0,0}{3.379}}\kern-0.35em} & {\kern-0.35em\colorbox[rgb]{0.7796078562736511,0.9174932767363155,0.8903652506716111}{\textcolor[rgb]{0,0,0}{2.073}}\kern-0.35em} & {\kern-0.35em\colorbox[rgb]{0.7419607996940613,0.9027297251364764,0.8689581008518443}{\textcolor[rgb]{0,0,0}{2.191}}\kern-0.35em} & {\kern-0.35em\colorbox[rgb]{0.7796078562736511,0.9174932767363155,0.8903652506716111}{\textcolor[rgb]{0,0,0}{2.076}}\kern-0.35em} & {\kern-0.35em\colorbox[rgb]{0.8084582966916701,0.9285351830370286,0.9083890865830815}{\textcolor[rgb]{0,0,0}{\kern-0.17em\textbf{1.963}\kern-0.17em}}\kern-0.35em}\\
 & \code{grammar.lsp} & \num{4} & {\kern-0.35em\colorbox[rgb]{0.3021299618131974,0.7078816014177659,0.5214609849686715}{\textcolor[rgb]{0,0,0}{3.666}}\kern-0.35em} & {\kern-0.35em\colorbox[rgb]{0.2452133934287464,0.6710496159160838,0.4469204183887033}{\textcolor[rgb]{1,1,1}{3.922}}\kern-0.35em} & {\kern-0.35em\colorbox[rgb]{0.31123415196643156,0.7128027852843789,0.5327797097318313}{\textcolor[rgb]{0,0,0}{3.633}}\kern-0.35em} & {\kern-0.35em\colorbox[rgb]{0.6729411959648133,0.8756632138701046,0.8297116595156052}{\textcolor[rgb]{0,0,0}{2.408}}\kern-0.35em} & {\kern-0.35em\colorbox[rgb]{0.5913725720901115,0.8433371873462902,0.7819761760094587}{\textcolor[rgb]{0,0,0}{2.657}}\kern-0.35em} & {\kern-0.35em\colorbox[rgb]{0.6854902148246766,0.8805843977367177,0.8368473761221942}{\textcolor[rgb]{0,0,0}{2.371}}\kern-0.35em} & {\kern-0.35em\colorbox[rgb]{0.7043137431144715,0.8879661735366372,0.8475509510320776}{\textcolor[rgb]{0,0,0}{\kern-0.17em\textbf{2.307}\kern-0.17em}}\kern-0.35em}\\
 & \code{lcet10.txt} & \num{417} & {\kern-0.35em\colorbox[rgb]{0.4972549166165146,0.8027374204467325,0.7136947498602025}{\textcolor[rgb]{0,0,0}{2.951}}\kern-0.35em} & {\kern-0.35em\colorbox[rgb]{0.4972549166165146,0.8027374204467325,0.7136947498602025}{\textcolor[rgb]{0,0,0}{2.953}}\kern-0.35em} & {\kern-0.35em\colorbox[rgb]{0.4972549166165146,0.8027374204467325,0.7136947498602025}{\textcolor[rgb]{0,0,0}{2.950}}\kern-0.35em} & {\kern-0.35em\colorbox[rgb]{0.8115340366083033,0.9296424494070166,0.9107266489197227}{\textcolor[rgb]{0,0,0}{1.946}}\kern-0.35em} & {\kern-0.35em\colorbox[rgb]{0.8146097765249365,0.9307497157770045,0.9130642112563638}{\textcolor[rgb]{0,0,0}{1.934}}\kern-0.35em} & {\kern-0.35em\colorbox[rgb]{0.8146097765249365,0.9307497157770045,0.9130642112563638}{\textcolor[rgb]{0,0,0}{1.931}}\kern-0.35em} & {\kern-0.35em\colorbox[rgb]{0.8207612563582027,0.9329642485169803,0.9177393359296462}{\textcolor[rgb]{0,0,0}{\kern-0.17em\textbf{1.897}\kern-0.17em}}\kern-0.35em}\\
 & \code{plrabn12.txt} & \num{471} & {\kern-0.35em\colorbox[rgb]{0.4219607922376371,0.7702576069270863,0.6590696089407977}{\textcolor[rgb]{0,0,0}{3.184}}\kern-0.35em} & {\kern-0.35em\colorbox[rgb]{0.4219607922376371,0.7702576069270863,0.6590696089407977}{\textcolor[rgb]{0,0,0}{3.186}}\kern-0.35em} & {\kern-0.35em\colorbox[rgb]{0.4219607922376371,0.7702576069270863,0.6590696089407977}{\textcolor[rgb]{0,0,0}{3.184}}\kern-0.35em} & {\kern-0.35em\colorbox[rgb]{0.6854902148246766,0.8805843977367177,0.8368473761221942}{\textcolor[rgb]{0,0,0}{2.364}}\kern-0.35em} & {\kern-0.35em\colorbox[rgb]{0.7043137431144715,0.8879661735366372,0.8475509510320776}{\textcolor[rgb]{0,0,0}{2.317}}\kern-0.35em} & {\kern-0.35em\colorbox[rgb]{0.7043137431144715,0.8879661735366372,0.8475509510320776}{\textcolor[rgb]{0,0,0}{2.314}}\kern-0.35em} & {\kern-0.35em\colorbox[rgb]{0.729411780834198,0.8978085412698633,0.8618223842452554}{\textcolor[rgb]{0,0,0}{\kern-0.17em\textbf{2.238}\kern-0.17em}}\kern-0.35em}\\
 & \code{xargs.1} & \num{4} & {\kern-0.35em\colorbox[rgb]{0.1972318446519328,0.6150711494333604,0.36855056531289043}{\textcolor[rgb]{1,1,1}{4.171}}\kern-0.35em} & {\kern-0.35em\colorbox[rgb]{0.15294118424256642,0.5633987188339233,0.29620916247367857}{\textcolor[rgb]{1,1,1}{4.391}}\kern-0.35em} & {\kern-0.35em\colorbox[rgb]{0.20092273301938002,0.6193771853166468,0.37457901554949147}{\textcolor[rgb]{1,1,1}{4.147}}\kern-0.35em} & {\kern-0.35em\colorbox[rgb]{0.48470589588670177,0.7973241181934582,0.7045905597069685}{\textcolor[rgb]{0,0,0}{2.992}}\kern-0.35em} & {\kern-0.35em\colorbox[rgb]{0.4219607922376371,0.7702576069270863,0.6590696089407977}{\textcolor[rgb]{0,0,0}{3.185}}\kern-0.35em} & {\kern-0.35em\colorbox[rgb]{0.4972549166165146,0.8027374204467325,0.7136947498602025}{\textcolor[rgb]{0,0,0}{2.941}}\kern-0.35em} & {\kern-0.35em\colorbox[rgb]{0.5223529580761405,0.8135640249532812,0.7319031301666709}{\textcolor[rgb]{0,0,0}{\kern-0.17em\textbf{2.869}\kern-0.17em}}\kern-0.35em}\\
\addlinespace[0em]\midrule\addlinespace[0.1em]
\multirow{11}{*}{\hspace{-2em}\rotatebox{65}{\emph{\utfeight}~}}
 & \code{ara-tabula.txt} & \num{2276} & {\kern-0.35em\colorbox[rgb]{0.8515186555245343,0.9440369122168597,0.9411149592960582}{\textcolor[rgb]{0,0,0}{1.692}}\kern-0.35em} & {\kern-0.35em\colorbox[rgb]{0.9044060033910415,0.963260286695817,0.9778854299994076}{\textcolor[rgb]{0,0,0}{1.343}}\kern-0.35em} & {\kern-0.35em\colorbox[rgb]{0.9044060033910415,0.963260286695817,0.9778854299994076}{\textcolor[rgb]{0,0,0}{1.342}}\kern-0.35em} & {\kern-0.35em\colorbox[rgb]{0.9331949290107279,0.9744559799923617,0.9842829690260045}{\textcolor[rgb]{0,0,0}{1.104}}\kern-0.35em} & {\kern-0.35em\colorbox[rgb]{0.9531257236705107,0.9822068445822771,0.9887120345059563}{\textcolor[rgb]{0,0,0}{0.929}}\kern-0.35em} & {\kern-0.35em\colorbox[rgb]{0.9531257236705107,0.9822068445822771,0.9887120345059563}{\textcolor[rgb]{0,0,0}{\kern-0.17em\textbf{0.928}\kern-0.17em}}\kern-0.35em} & {\kern-0.35em\colorbox[rgb]{0.913264134350945,0.9667051154024461,0.9798539035460528}{\textcolor[rgb]{0,0,0}{1.268}}\kern-0.35em}\\
 & \code{ben-kobita.txt} & \num{420} & {\kern-0.35em\colorbox[rgb]{0.8330642160247355,0.937393313996932,0.927089585276211}{\textcolor[rgb]{0,0,0}{1.804}}\kern-0.35em} & {\kern-0.35em\colorbox[rgb]{0.8945790143573985,0.9595386413966909,0.973840832009035}{\textcolor[rgb]{0,0,0}{1.427}}\kern-0.35em} & {\kern-0.35em\colorbox[rgb]{0.8945790143573985,0.9595386413966909,0.973840832009035}{\textcolor[rgb]{0,0,0}{1.423}}\kern-0.35em} & {\kern-0.35em\colorbox[rgb]{0.9199077325708725,0.9692887369324179,0.9813302587060367}{\textcolor[rgb]{0,0,0}{1.222}}\kern-0.35em} & {\kern-0.35em\colorbox[rgb]{0.9331949290107279,0.9744559799923617,0.9842829690260045}{\textcolor[rgb]{0,0,0}{1.097}}\kern-0.35em} & {\kern-0.35em\colorbox[rgb]{0.9331949290107279,0.9744559799923617,0.9842829690260045}{\textcolor[rgb]{0,0,0}{\kern-0.17em\textbf{1.093}\kern-0.17em}}\kern-0.35em} & {\kern-0.35em\colorbox[rgb]{0.8999769379110898,0.9615378723425023,0.976901193226085}{\textcolor[rgb]{0,0,0}{1.380}}\kern-0.35em}\\
 & \code{hin-baital.txt} & \num{210} & {\kern-0.35em\colorbox[rgb]{0.8515186555245343,0.9440369122168597,0.9411149592960582}{\textcolor[rgb]{0,0,0}{1.692}}\kern-0.35em} & {\kern-0.35em\colorbox[rgb]{0.9154786670909209,0.9675663225791034,0.9803460219327141}{\textcolor[rgb]{0,0,0}{1.246}}\kern-0.35em} & {\kern-0.35em\colorbox[rgb]{0.9176931998308967,0.9684275297557606,0.9808381403193754}{\textcolor[rgb]{0,0,0}{1.241}}\kern-0.35em} & {\kern-0.35em\colorbox[rgb]{0.9442675927106072,0.978762015875648,0.986743560959311}{\textcolor[rgb]{0,0,0}{0.995}}\kern-0.35em} & {\kern-0.35em\colorbox[rgb]{0.966412920110366,0.9873740876422209,0.9916647448259241}{\textcolor[rgb]{0,0,0}{0.815}}\kern-0.35em} & {\kern-0.35em\colorbox[rgb]{0.966412920110366,0.9873740876422209,0.9916647448259241}{\textcolor[rgb]{0,0,0}{\kern-0.17em\textbf{0.810}\kern-0.17em}}\kern-0.35em} & {\kern-0.35em\colorbox[rgb]{0.9199077325708725,0.9692887369324179,0.9813302587060367}{\textcolor[rgb]{0,0,0}{1.221}}\kern-0.35em}\\
 & \code{jav-tuban.txt} & \num{138} & {\kern-0.35em\colorbox[rgb]{0.49098040625160816,0.8000307693200953,0.7091426547835855}{\textcolor[rgb]{0,0,0}{2.962}}\kern-0.35em} & {\kern-0.35em\colorbox[rgb]{0.5035294269814211,0.8054440715733696,0.7182468449368197}{\textcolor[rgb]{0,0,0}{2.935}}\kern-0.35em} & {\kern-0.35em\colorbox[rgb]{0.5035294269814211,0.8054440715733696,0.7182468449368197}{\textcolor[rgb]{0,0,0}{2.926}}\kern-0.35em} & {\kern-0.35em\colorbox[rgb]{0.7482353091239929,0.9051903170697829,0.8725259591551388}{\textcolor[rgb]{0,0,0}{2.187}}\kern-0.35em} & {\kern-0.35em\colorbox[rgb]{0.7482353091239929,0.9051903170697829,0.8725259591551388}{\textcolor[rgb]{0,0,0}{2.177}}\kern-0.35em} & {\kern-0.35em\colorbox[rgb]{0.7545098185539245,0.9076509090030894,0.8760938174584333}{\textcolor[rgb]{0,0,0}{2.168}}\kern-0.35em} & {\kern-0.35em\colorbox[rgb]{0.7607843279838562,0.910111500936396,0.8796616757617277}{\textcolor[rgb]{0,0,0}{\kern-0.17em\textbf{2.140}\kern-0.17em}}\kern-0.35em}\\
 & \code{jpn-yujo.txt} & \num{199} & {\kern-0.35em\colorbox[rgb]{0.5913725720901115,0.8433371873462902,0.7819761760094587}{\textcolor[rgb]{0,0,0}{2.658}}\kern-0.35em} & {\kern-0.35em\colorbox[rgb]{0.7921568751335144,0.9224144606029286,0.8975009672782001}{\textcolor[rgb]{0,0,0}{2.038}}\kern-0.35em} & {\kern-0.35em\colorbox[rgb]{0.8023068168584038,0.9263206502970527,0.903713961909799}{\textcolor[rgb]{0,0,0}{2.002}}\kern-0.35em} & {\kern-0.35em\colorbox[rgb]{0.8545943954411674,0.9451441785868476,0.9434525216326994}{\textcolor[rgb]{0,0,0}{1.684}}\kern-0.35em} & {\kern-0.35em\colorbox[rgb]{0.8638216151910669,0.9484659776968114,0.950465208642623}{\textcolor[rgb]{0,0,0}{1.621}}\kern-0.35em} & {\kern-0.35em\colorbox[rgb]{0.8699730950243333,0.9506805104367874,0.9551403333159054}{\textcolor[rgb]{0,0,0}{\kern-0.17em\textbf{1.585}\kern-0.17em}}\kern-0.35em} & {\kern-0.35em\colorbox[rgb]{0.8453671756912681,0.9418223794768839,0.9364398346227758}{\textcolor[rgb]{0,0,0}{1.736}}\kern-0.35em}\\
 & \code{lah-wiki.txt} & \num{121} & {\kern-0.35em\colorbox[rgb]{0.597647082455018,0.8460438384729273,0.7865282710860757}{\textcolor[rgb]{0,0,0}{2.638}}\kern-0.35em} & {\kern-0.35em\colorbox[rgb]{0.729411780834198,0.8978085412698633,0.8618223842452554}{\textcolor[rgb]{0,0,0}{2.232}}\kern-0.35em} & {\kern-0.35em\colorbox[rgb]{0.7356862902641297,0.9002691332031699,0.8653902425485499}{\textcolor[rgb]{0,0,0}{2.218}}\kern-0.35em} & {\kern-0.35em\colorbox[rgb]{0.8422914357746348,0.9407151131068959,0.9341022722861346}{\textcolor[rgb]{0,0,0}{1.748}}\kern-0.35em} & {\kern-0.35em\colorbox[rgb]{0.8545943954411674,0.9451441785868476,0.9434525216326994}{\textcolor[rgb]{0,0,0}{1.686}}\kern-0.35em} & {\kern-0.35em\colorbox[rgb]{0.8545943954411674,0.9451441785868476,0.9434525216326994}{\textcolor[rgb]{0,0,0}{\kern-0.17em\textbf{1.672}\kern-0.17em}}\kern-0.35em} & {\kern-0.35em\colorbox[rgb]{0.8269127361914691,0.9351787812569562,0.9224144606029286}{\textcolor[rgb]{0,0,0}{1.847}}\kern-0.35em}\\
 & \code{por-noites.txt} & \num{113} & {\kern-0.35em\colorbox[rgb]{0.31123415196643156,0.7128027852843789,0.5327797097318313}{\textcolor[rgb]{0,0,0}{3.625}}\kern-0.35em} & {\kern-0.35em\colorbox[rgb]{0.31578624704304864,0.7152633772176855,0.5384390721134111}{\textcolor[rgb]{0,0,0}{3.607}}\kern-0.35em} & {\kern-0.35em\colorbox[rgb]{0.3203383421196657,0.717723969150992,0.544098434494991}{\textcolor[rgb]{0,0,0}{3.596}}\kern-0.35em} & {\kern-0.35em\colorbox[rgb]{0.5349019788059535,0.8189773272065556,0.741007320319905}{\textcolor[rgb]{0,0,0}{2.834}}\kern-0.35em} & {\kern-0.35em\colorbox[rgb]{0.5474509995357663,0.8243906294598299,0.7501115104731392}{\textcolor[rgb]{0,0,0}{2.801}}\kern-0.35em} & {\kern-0.35em\colorbox[rgb]{0.5474509995357663,0.8243906294598299,0.7501115104731392}{\textcolor[rgb]{0,0,0}{2.789}}\kern-0.35em} & {\kern-0.35em\colorbox[rgb]{0.5913725720901115,0.8433371873462902,0.7819761760094587}{\textcolor[rgb]{0,0,0}{\kern-0.17em\textbf{2.663}\kern-0.17em}}\kern-0.35em}\\
 & \code{rus-mosco.txt} & \num{80} & {\kern-0.35em\colorbox[rgb]{0.5349019788059535,0.8189773272065556,0.741007320319905}{\textcolor[rgb]{0,0,0}{2.839}}\kern-0.35em} & {\kern-0.35em\colorbox[rgb]{0.6666666865348816,0.8732026219367981,0.8261438012123108}{\textcolor[rgb]{0,0,0}{2.438}}\kern-0.35em} & {\kern-0.35em\colorbox[rgb]{0.6729411959648133,0.8756632138701046,0.8297116595156052}{\textcolor[rgb]{0,0,0}{2.415}}\kern-0.35em} & {\kern-0.35em\colorbox[rgb]{0.8269127361914691,0.9351787812569562,0.9224144606029286}{\textcolor[rgb]{0,0,0}{1.846}}\kern-0.35em} & {\kern-0.35em\colorbox[rgb]{0.8361399559413686,0.93850058036692,0.9294271476128522}{\textcolor[rgb]{0,0,0}{1.796}}\kern-0.35em} & {\kern-0.35em\colorbox[rgb]{0.8392156958580017,0.939607846736908,0.9317647099494935}{\textcolor[rgb]{0,0,0}{\kern-0.17em\textbf{1.772}\kern-0.17em}}\kern-0.35em} & {\kern-0.35em\colorbox[rgb]{0.8207612563582027,0.9329642485169803,0.9177393359296462}{\textcolor[rgb]{0,0,0}{1.895}}\kern-0.35em}\\
 & \code{spa-trans.txt} & \num{304} & {\kern-0.35em\colorbox[rgb]{0.4407843233323565,0.7783775603069979,0.6727258941706489}{\textcolor[rgb]{0,0,0}{3.116}}\kern-0.35em} & {\kern-0.35em\colorbox[rgb]{0.4533333440621694,0.7837908625602722,0.681830084323883}{\textcolor[rgb]{0,0,0}{3.083}}\kern-0.35em} & {\kern-0.35em\colorbox[rgb]{0.4533333440621694,0.7837908625602722,0.681830084323883}{\textcolor[rgb]{0,0,0}{3.079}}\kern-0.35em} & {\kern-0.35em\colorbox[rgb]{0.7168627619743347,0.8928873574032503,0.8546866676386665}{\textcolor[rgb]{0,0,0}{2.266}}\kern-0.35em} & {\kern-0.35em\colorbox[rgb]{0.729411780834198,0.8978085412698633,0.8618223842452554}{\textcolor[rgb]{0,0,0}{2.236}}\kern-0.35em} & {\kern-0.35em\colorbox[rgb]{0.729411780834198,0.8978085412698633,0.8618223842452554}{\textcolor[rgb]{0,0,0}{2.232}}\kern-0.35em} & {\kern-0.35em\colorbox[rgb]{0.7545098185539245,0.9076509090030894,0.8760938174584333}{\textcolor[rgb]{0,0,0}{\kern-0.17em\textbf{2.169}\kern-0.17em}}\kern-0.35em}\\
 & \code{zho-you.txt} & \num{63} & {\kern-0.35em\colorbox[rgb]{0.0823529452085495,0.4980392336845398,0.23137255907058715}{\textcolor[rgb]{1,1,1}{4.730}}\kern-0.35em} & {\kern-0.35em\colorbox[rgb]{0.15294118424256642,0.5633987188339233,0.29620916247367857}{\textcolor[rgb]{1,1,1}{4.391}}\kern-0.35em} & {\kern-0.35em\colorbox[rgb]{0.20092273301938002,0.6193771853166468,0.37457901554949147}{\textcolor[rgb]{1,1,1}{4.146}}\kern-0.35em} & {\kern-0.35em\colorbox[rgb]{0.274817391353495,0.6931180498179268,0.4875048106791926}{\textcolor[rgb]{1,1,1}{3.787}}\kern-0.35em} & {\kern-0.35em\colorbox[rgb]{0.31123415196643156,0.7128027852843789,0.5327797097318313}{\textcolor[rgb]{0,0,0}{3.628}}\kern-0.35em} & {\kern-0.35em\colorbox[rgb]{0.3704113879624535,0.7447904804173637,0.606351420692369}{\textcolor[rgb]{0,0,0}{\kern-0.17em\textbf{3.383}\kern-0.17em}}\kern-0.35em} & {\kern-0.35em\colorbox[rgb]{0.32489043719628274,0.7201845610842985,0.5497577968765708}{\textcolor[rgb]{0,0,0}{3.563}}\kern-0.35em}\\
 & \code{mix-sake.txt} & \num{251} & {\kern-0.35em\colorbox[rgb]{0.34765091257936814,0.7324875207508311,0.5780546087844699}{\textcolor[rgb]{0,0,0}{3.476}}\kern-0.35em} & {\kern-0.35em\colorbox[rgb]{0.45960785442707586,0.7864975136869095,0.6863821794005002}{\textcolor[rgb]{0,0,0}{3.060}}\kern-0.35em} & {\kern-0.35em\colorbox[rgb]{0.47215687515688876,0.7919108159401838,0.6954863695537343}{\textcolor[rgb]{0,0,0}{3.026}}\kern-0.35em} & {\kern-0.35em\colorbox[rgb]{0.7733333468437195,0.915032684803009,0.8867973923683167}{\textcolor[rgb]{0,0,0}{2.102}}\kern-0.35em} & {\kern-0.35em\colorbox[rgb]{0.7858823657035828,0.9199538686696221,0.8939331089749056}{\textcolor[rgb]{0,0,0}{2.060}}\kern-0.35em} & {\kern-0.35em\colorbox[rgb]{0.7984313845634461,0.9248750525362351,0.9010688255814945}{\textcolor[rgb]{0,0,0}{\kern-0.17em\textbf{2.026}\kern-0.17em}}\kern-0.35em} & {\kern-0.35em\colorbox[rgb]{0.7796078562736511,0.9174932767363155,0.8903652506716111}{\textcolor[rgb]{0,0,0}{2.076}}\kern-0.35em}\\
\addlinespace[0em]\midrule\addlinespace[0.1em]
\multirow{3}{*}{\hspace{-2em}\rotatebox{65}{\emph{Binary}~}}
 & \code{kennedy.xls} & \num{1006} & {\kern-0.35em\colorbox[rgb]{0.6729411959648133,0.8756632138701046,0.8297116595156052}{\textcolor[rgb]{0,0,0}{2.413}}\kern-0.35em} & {\kern-0.35em\colorbox[rgb]{0.6729411959648133,0.8756632138701046,0.8297116595156052}{\textcolor[rgb]{0,0,0}{2.418}}\kern-0.35em} & {\kern-0.35em\colorbox[rgb]{0.6729411959648133,0.8756632138701046,0.8297116595156052}{\textcolor[rgb]{0,0,0}{2.414}}\kern-0.35em} & {\kern-0.35em\colorbox[rgb]{0.8699730950243333,0.9506805104367874,0.9551403333159054}{\textcolor[rgb]{0,0,0}{1.586}}\kern-0.35em} & {\kern-0.35em\colorbox[rgb]{0.8884275345241323,0.9573241086567149,0.9691657073357526}{\textcolor[rgb]{0,0,0}{1.474}}\kern-0.35em} & {\kern-0.35em\colorbox[rgb]{0.8884275345241323,0.9573241086567149,0.9691657073357526}{\textcolor[rgb]{0,0,0}{1.470}}\kern-0.35em} & {\kern-0.35em\colorbox[rgb]{0.9531257236705107,0.9822068445822771,0.9887120345059563}{\textcolor[rgb]{0,0,0}{\kern-0.17em\textbf{0.919}\kern-0.17em}}\kern-0.35em}\\
 & \code{ptt5} & \num{501} & {\kern-0.35em\colorbox[rgb]{0.9509111909305348,0.9813456374056199,0.988219916119295}{\textcolor[rgb]{0,0,0}{0.936}}\kern-0.35em} & {\kern-0.35em\colorbox[rgb]{0.9509111909305348,0.9813456374056199,0.988219916119295}{\textcolor[rgb]{0,0,0}{0.946}}\kern-0.35em} & {\kern-0.35em\colorbox[rgb]{0.9509111909305348,0.9813456374056199,0.988219916119295}{\textcolor[rgb]{0,0,0}{0.941}}\kern-0.35em} & {\kern-0.35em\colorbox[rgb]{0.9641983873703901,0.9865128804655636,0.9911726264392628}{\textcolor[rgb]{0,0,0}{0.824}}\kern-0.35em} & {\kern-0.35em\colorbox[rgb]{0.9641983873703901,0.9865128804655636,0.9911726264392628}{\textcolor[rgb]{0,0,0}{0.821}}\kern-0.35em} & {\kern-0.35em\colorbox[rgb]{0.966412920110366,0.9873740876422209,0.9916647448259241}{\textcolor[rgb]{0,0,0}{0.816}}\kern-0.35em} & {\kern-0.35em\colorbox[rgb]{0.9686274528503418,0.9882352948188782,0.9921568632125854}{\textcolor[rgb]{0,0,0}{\kern-0.17em\textbf{0.781}\kern-0.17em}}\kern-0.35em}\\
 & \code{sum} & \num{37} & {\kern-0.35em\colorbox[rgb]{0.219377174856616,0.6409073647330789,0.40472126673249637}{\textcolor[rgb]{1,1,1}{4.051}}\kern-0.35em} & {\kern-0.35em\colorbox[rgb]{0.1972318446519328,0.6150711494333604,0.36855056531289043}{\textcolor[rgb]{1,1,1}{4.161}}\kern-0.35em} & {\kern-0.35em\colorbox[rgb]{0.21568628648916882,0.6366013288497925,0.3986928164958954}{\textcolor[rgb]{1,1,1}{4.067}}\kern-0.35em} & {\kern-0.35em\colorbox[rgb]{0.5662745306304857,0.8325105828397414,0.7637677957029904}{\textcolor[rgb]{0,0,0}{2.734}}\kern-0.35em} & {\kern-0.35em\colorbox[rgb]{0.5349019788059535,0.8189773272065556,0.741007320319905}{\textcolor[rgb]{0,0,0}{2.831}}\kern-0.35em} & {\kern-0.35em\colorbox[rgb]{0.5662745306304857,0.8325105828397414,0.7637677957029904}{\textcolor[rgb]{0,0,0}{2.737}}\kern-0.35em} & {\kern-0.35em\colorbox[rgb]{0.6541176676750183,0.8682814380701851,0.8190080846057218}{\textcolor[rgb]{0,0,0}{\kern-0.17em\textbf{2.469}\kern-0.17em}}\kern-0.35em}\\
\bottomrule
\end{tabular}

\caption[Compression effectiveness]{Compression effectiveness in \bitsperbyte. Column key: first character indicates whether compressor is based on \lzw (\humble{L}) or \ppm (\humble{P}); remaining characters specify whether base model is uniform byte (\uniformbyte), uniform token (\uniformtoken) or P\'{o}lya token (\polyatoken). For comparison, we also include \ppmii, a \ppm variant by Shkarin~\cite{shkarin:2006a}. All figures are given to 3 decimal places. Each cell is shaded to indicate how good the compression rate is relative to other compressors in the table. The best compressor in each row is in \textbf{bold}.}
\label{table:eval:per-file}
\vspace*{0.5em}
\centerline{\raisebox{-0.25em}{\textsf{worse} $\leftarrow$}\hspace{0.5em}{\colorbox[rgb]{0.0823529452085495,0.4980392336845398,0.23137255907058715}{\textcolor[rgb]{1,1,1}{\kern-0.25em\hspace*{0.006999999999999999\textwidth}\kern-0.25em}}}{\colorbox[rgb]{0.09096501791009716,0.505421011354409,0.2375240398388283}{\textcolor[rgb]{1,1,1}{\kern-0.25em\hspace*{0.006999999999999999\textwidth}\kern-0.25em}}}{\colorbox[rgb]{0.09957709061164483,0.5128027890242782,0.24367552060706943}{\textcolor[rgb]{1,1,1}{\kern-0.25em\hspace*{0.006999999999999999\textwidth}\kern-0.25em}}}{\colorbox[rgb]{0.10818916331319248,0.5201845666941475,0.24982700137531055}{\textcolor[rgb]{1,1,1}{\kern-0.25em\hspace*{0.006999999999999999\textwidth}\kern-0.25em}}}{\colorbox[rgb]{0.11680123601474027,0.5275663443640167,0.2559784821435518}{\textcolor[rgb]{1,1,1}{\kern-0.25em\hspace*{0.006999999999999999\textwidth}\kern-0.25em}}}{\colorbox[rgb]{0.1254133087162878,0.5349481220338859,0.2621299629117928}{\textcolor[rgb]{1,1,1}{\kern-0.25em\hspace*{0.006999999999999999\textwidth}\kern-0.25em}}}{\colorbox[rgb]{0.13402538141783546,0.5423298997037551,0.268281443680034}{\textcolor[rgb]{1,1,1}{\kern-0.25em\hspace*{0.006999999999999999\textwidth}\kern-0.25em}}}{\colorbox[rgb]{0.14186851914022483,0.5504806111840641,0.27812381176387563}{\textcolor[rgb]{1,1,1}{\kern-0.25em\hspace*{0.006999999999999999\textwidth}\kern-0.25em}}}{\colorbox[rgb]{0.14925029587511923,0.5590926829506369,0.2901807122370776}{\textcolor[rgb]{1,1,1}{\kern-0.25em\hspace*{0.006999999999999999\textwidth}\kern-0.25em}}}{\colorbox[rgb]{0.1566320726100136,0.5677047547172097,0.3022376127102796}{\textcolor[rgb]{1,1,1}{\kern-0.25em\hspace*{0.006999999999999999\textwidth}\kern-0.25em}}}{\colorbox[rgb]{0.16401384934490804,0.5763168264837826,0.31429451318348156}{\textcolor[rgb]{1,1,1}{\kern-0.25em\hspace*{0.006999999999999999\textwidth}\kern-0.25em}}}{\colorbox[rgb]{0.1713956260798024,0.5849288982503554,0.3263514136566835}{\textcolor[rgb]{1,1,1}{\kern-0.25em\hspace*{0.006999999999999999\textwidth}\kern-0.25em}}}{\colorbox[rgb]{0.17877740281469692,0.5935409700169284,0.3384083141298857}{\textcolor[rgb]{1,1,1}{\kern-0.25em\hspace*{0.006999999999999999\textwidth}\kern-0.25em}}}{\colorbox[rgb]{0.1861591795495912,0.6021530417835012,0.3504652146030875}{\textcolor[rgb]{1,1,1}{\kern-0.25em\hspace*{0.006999999999999999\textwidth}\kern-0.25em}}}{\colorbox[rgb]{0.19354095628448562,0.610765113550074,0.36252211507628945}{\textcolor[rgb]{1,1,1}{\kern-0.25em\hspace*{0.006999999999999999\textwidth}\kern-0.25em}}}{\colorbox[rgb]{0.20092273301938002,0.6193771853166468,0.37457901554949147}{\textcolor[rgb]{1,1,1}{\kern-0.25em\hspace*{0.006999999999999999\textwidth}\kern-0.25em}}}{\colorbox[rgb]{0.2083045097542744,0.6279892570832196,0.3866359160226934}{\textcolor[rgb]{1,1,1}{\kern-0.25em\hspace*{0.006999999999999999\textwidth}\kern-0.25em}}}{\colorbox[rgb]{0.219377174856616,0.6409073647330789,0.40472126673249637}{\textcolor[rgb]{1,1,1}{\kern-0.25em\hspace*{0.006999999999999999\textwidth}\kern-0.25em}}}{\colorbox[rgb]{0.22675895159151038,0.6495194364996517,0.4167781672056983}{\textcolor[rgb]{1,1,1}{\kern-0.25em\hspace*{0.006999999999999999\textwidth}\kern-0.25em}}}{\colorbox[rgb]{0.23414072832640478,0.6581315082662246,0.42883506767890034}{\textcolor[rgb]{1,1,1}{\kern-0.25em\hspace*{0.006999999999999999\textwidth}\kern-0.25em}}}{\colorbox[rgb]{0.2415225050612992,0.6667435800327974,0.4408919681521023}{\textcolor[rgb]{1,1,1}{\kern-0.25em\hspace*{0.006999999999999999\textwidth}\kern-0.25em}}}{\colorbox[rgb]{0.2489042817961936,0.6753556517993703,0.45294886862530426}{\textcolor[rgb]{1,1,1}{\kern-0.25em\hspace*{0.006999999999999999\textwidth}\kern-0.25em}}}{\colorbox[rgb]{0.2566090110470267,0.6832756820847006,0.4648673611528733}{\textcolor[rgb]{1,1,1}{\kern-0.25em\hspace*{0.006999999999999999\textwidth}\kern-0.25em}}}{\colorbox[rgb]{0.2657132012002608,0.6881968659513137,0.47618608591603295}{\textcolor[rgb]{1,1,1}{\kern-0.25em\hspace*{0.006999999999999999\textwidth}\kern-0.25em}}}{\colorbox[rgb]{0.274817391353495,0.6931180498179268,0.4875048106791926}{\textcolor[rgb]{1,1,1}{\kern-0.25em\hspace*{0.006999999999999999\textwidth}\kern-0.25em}}}{\colorbox[rgb]{0.28392158150672914,0.6980392336845398,0.49882353544235225}{\textcolor[rgb]{1,1,1}{\kern-0.25em\hspace*{0.006999999999999999\textwidth}\kern-0.25em}}}{\colorbox[rgb]{0.29302577165996324,0.7029604175511528,0.5101422602055119}{\textcolor[rgb]{0,0,0}{\kern-0.25em\hspace*{0.006999999999999999\textwidth}\kern-0.25em}}}{\colorbox[rgb]{0.3021299618131974,0.7078816014177659,0.5214609849686715}{\textcolor[rgb]{0,0,0}{\kern-0.25em\hspace*{0.006999999999999999\textwidth}\kern-0.25em}}}{\colorbox[rgb]{0.31123415196643156,0.7128027852843789,0.5327797097318313}{\textcolor[rgb]{0,0,0}{\kern-0.25em\hspace*{0.006999999999999999\textwidth}\kern-0.25em}}}{\colorbox[rgb]{0.3203383421196657,0.717723969150992,0.544098434494991}{\textcolor[rgb]{0,0,0}{\kern-0.25em\hspace*{0.006999999999999999\textwidth}\kern-0.25em}}}{\colorbox[rgb]{0.3294425322728999,0.722645153017605,0.5554171592581506}{\textcolor[rgb]{0,0,0}{\kern-0.25em\hspace*{0.006999999999999999\textwidth}\kern-0.25em}}}{\colorbox[rgb]{0.338546722426134,0.7275663368842181,0.5667358840213103}{\textcolor[rgb]{0,0,0}{\kern-0.25em\hspace*{0.006999999999999999\textwidth}\kern-0.25em}}}{\colorbox[rgb]{0.34765091257936814,0.7324875207508311,0.5780546087844699}{\textcolor[rgb]{0,0,0}{\kern-0.25em\hspace*{0.006999999999999999\textwidth}\kern-0.25em}}}{\colorbox[rgb]{0.3567551027326023,0.7374087046174442,0.5893733335476296}{\textcolor[rgb]{0,0,0}{\kern-0.25em\hspace*{0.006999999999999999\textwidth}\kern-0.25em}}}{\colorbox[rgb]{0.3658592928858364,0.7423298884840572,0.6006920583107892}{\textcolor[rgb]{0,0,0}{\kern-0.25em\hspace*{0.006999999999999999\textwidth}\kern-0.25em}}}{\colorbox[rgb]{0.37496348303907057,0.7472510723506703,0.6120107830739488}{\textcolor[rgb]{0,0,0}{\kern-0.25em\hspace*{0.006999999999999999\textwidth}\kern-0.25em}}}{\colorbox[rgb]{0.3840676731923047,0.7521722562172833,0.6233295078371085}{\textcolor[rgb]{0,0,0}{\kern-0.25em\hspace*{0.006999999999999999\textwidth}\kern-0.25em}}}{\colorbox[rgb]{0.3931718633455389,0.7570934400838965,0.6346482326002681}{\textcolor[rgb]{0,0,0}{\kern-0.25em\hspace*{0.006999999999999999\textwidth}\kern-0.25em}}}{\colorbox[rgb]{0.40941177150782415,0.7648443046738119,0.6499654187875635}{\textcolor[rgb]{0,0,0}{\kern-0.25em\hspace*{0.006999999999999999\textwidth}\kern-0.25em}}}{\colorbox[rgb]{0.4219607922376371,0.7702576069270863,0.6590696089407977}{\textcolor[rgb]{0,0,0}{\kern-0.25em\hspace*{0.006999999999999999\textwidth}\kern-0.25em}}}{\colorbox[rgb]{0.4345098129674501,0.7756709091803607,0.668173799094032}{\textcolor[rgb]{0,0,0}{\kern-0.25em\hspace*{0.006999999999999999\textwidth}\kern-0.25em}}}{\colorbox[rgb]{0.44705883369726296,0.781084211433635,0.677277989247266}{\textcolor[rgb]{0,0,0}{\kern-0.25em\hspace*{0.006999999999999999\textwidth}\kern-0.25em}}}{\colorbox[rgb]{0.45960785442707586,0.7864975136869095,0.6863821794005002}{\textcolor[rgb]{0,0,0}{\kern-0.25em\hspace*{0.006999999999999999\textwidth}\kern-0.25em}}}{\colorbox[rgb]{0.47215687515688876,0.7919108159401838,0.6954863695537343}{\textcolor[rgb]{0,0,0}{\kern-0.25em\hspace*{0.006999999999999999\textwidth}\kern-0.25em}}}{\colorbox[rgb]{0.48470589588670177,0.7973241181934582,0.7045905597069685}{\textcolor[rgb]{0,0,0}{\kern-0.25em\hspace*{0.006999999999999999\textwidth}\kern-0.25em}}}{\colorbox[rgb]{0.4972549166165146,0.8027374204467325,0.7136947498602025}{\textcolor[rgb]{0,0,0}{\kern-0.25em\hspace*{0.006999999999999999\textwidth}\kern-0.25em}}}{\colorbox[rgb]{0.5098039373463276,0.8081507227000069,0.7227989400134367}{\textcolor[rgb]{0,0,0}{\kern-0.25em\hspace*{0.006999999999999999\textwidth}\kern-0.25em}}}{\colorbox[rgb]{0.5223529580761405,0.8135640249532812,0.7319031301666709}{\textcolor[rgb]{0,0,0}{\kern-0.25em\hspace*{0.006999999999999999\textwidth}\kern-0.25em}}}{\colorbox[rgb]{0.5349019788059535,0.8189773272065556,0.741007320319905}{\textcolor[rgb]{0,0,0}{\kern-0.25em\hspace*{0.006999999999999999\textwidth}\kern-0.25em}}}{\colorbox[rgb]{0.5474509995357663,0.8243906294598299,0.7501115104731392}{\textcolor[rgb]{0,0,0}{\kern-0.25em\hspace*{0.006999999999999999\textwidth}\kern-0.25em}}}{\colorbox[rgb]{0.5600000202655793,0.8298039317131043,0.7592157006263733}{\textcolor[rgb]{0,0,0}{\kern-0.25em\hspace*{0.006999999999999999\textwidth}\kern-0.25em}}}{\colorbox[rgb]{0.5725490409953922,0.8352172339663786,0.7683198907796074}{\textcolor[rgb]{0,0,0}{\kern-0.25em\hspace*{0.006999999999999999\textwidth}\kern-0.25em}}}{\colorbox[rgb]{0.5850980617252052,0.840630536219653,0.7774240809328417}{\textcolor[rgb]{0,0,0}{\kern-0.25em\hspace*{0.006999999999999999\textwidth}\kern-0.25em}}}{\colorbox[rgb]{0.597647082455018,0.8460438384729273,0.7865282710860757}{\textcolor[rgb]{0,0,0}{\kern-0.25em\hspace*{0.006999999999999999\textwidth}\kern-0.25em}}}{\colorbox[rgb]{0.6101961016654969,0.8510572945370394,0.7940330764826606}{\textcolor[rgb]{0,0,0}{\kern-0.25em\hspace*{0.006999999999999999\textwidth}\kern-0.25em}}}{\colorbox[rgb]{0.6227451205253601,0.8559784784036524,0.8011687930892496}{\textcolor[rgb]{0,0,0}{\kern-0.25em\hspace*{0.006999999999999999\textwidth}\kern-0.25em}}}{\colorbox[rgb]{0.6352941393852235,0.8608996622702655,0.8083045096958386}{\textcolor[rgb]{0,0,0}{\kern-0.25em\hspace*{0.006999999999999999\textwidth}\kern-0.25em}}}{\colorbox[rgb]{0.6478431582450866,0.8658208461368785,0.8154402263024274}{\textcolor[rgb]{0,0,0}{\kern-0.25em\hspace*{0.006999999999999999\textwidth}\kern-0.25em}}}{\colorbox[rgb]{0.6603921771049499,0.8707420300034916,0.8225759429090164}{\textcolor[rgb]{0,0,0}{\kern-0.25em\hspace*{0.006999999999999999\textwidth}\kern-0.25em}}}{\colorbox[rgb]{0.6792157053947449,0.8781238058034111,0.8332795178188998}{\textcolor[rgb]{0,0,0}{\kern-0.25em\hspace*{0.006999999999999999\textwidth}\kern-0.25em}}}{\colorbox[rgb]{0.6917647242546081,0.8830449896700242,0.8404152344254886}{\textcolor[rgb]{0,0,0}{\kern-0.25em\hspace*{0.006999999999999999\textwidth}\kern-0.25em}}}{\colorbox[rgb]{0.7043137431144715,0.8879661735366372,0.8475509510320776}{\textcolor[rgb]{0,0,0}{\kern-0.25em\hspace*{0.006999999999999999\textwidth}\kern-0.25em}}}{\colorbox[rgb]{0.7168627619743347,0.8928873574032503,0.8546866676386665}{\textcolor[rgb]{0,0,0}{\kern-0.25em\hspace*{0.006999999999999999\textwidth}\kern-0.25em}}}{\colorbox[rgb]{0.729411780834198,0.8978085412698633,0.8618223842452554}{\textcolor[rgb]{0,0,0}{\kern-0.25em\hspace*{0.006999999999999999\textwidth}\kern-0.25em}}}{\colorbox[rgb]{0.7419607996940613,0.9027297251364764,0.8689581008518443}{\textcolor[rgb]{0,0,0}{\kern-0.25em\hspace*{0.006999999999999999\textwidth}\kern-0.25em}}}{\colorbox[rgb]{0.7545098185539245,0.9076509090030894,0.8760938174584333}{\textcolor[rgb]{0,0,0}{\kern-0.25em\hspace*{0.006999999999999999\textwidth}\kern-0.25em}}}{\colorbox[rgb]{0.7670588374137879,0.9125720928697025,0.8832295340650221}{\textcolor[rgb]{0,0,0}{\kern-0.25em\hspace*{0.006999999999999999\textwidth}\kern-0.25em}}}{\colorbox[rgb]{0.7796078562736511,0.9174932767363155,0.8903652506716111}{\textcolor[rgb]{0,0,0}{\kern-0.25em\hspace*{0.006999999999999999\textwidth}\kern-0.25em}}}{\colorbox[rgb]{0.7921568751335144,0.9224144606029286,0.8975009672782001}{\textcolor[rgb]{0,0,0}{\kern-0.25em\hspace*{0.006999999999999999\textwidth}\kern-0.25em}}}{\colorbox[rgb]{0.8023068168584038,0.9263206502970527,0.903713961909799}{\textcolor[rgb]{0,0,0}{\kern-0.25em\hspace*{0.006999999999999999\textwidth}\kern-0.25em}}}{\colorbox[rgb]{0.8084582966916701,0.9285351830370286,0.9083890865830815}{\textcolor[rgb]{0,0,0}{\kern-0.25em\hspace*{0.006999999999999999\textwidth}\kern-0.25em}}}{\colorbox[rgb]{0.8146097765249365,0.9307497157770045,0.9130642112563638}{\textcolor[rgb]{0,0,0}{\kern-0.25em\hspace*{0.006999999999999999\textwidth}\kern-0.25em}}}{\colorbox[rgb]{0.8207612563582027,0.9329642485169803,0.9177393359296462}{\textcolor[rgb]{0,0,0}{\kern-0.25em\hspace*{0.006999999999999999\textwidth}\kern-0.25em}}}{\colorbox[rgb]{0.8269127361914691,0.9351787812569562,0.9224144606029286}{\textcolor[rgb]{0,0,0}{\kern-0.25em\hspace*{0.006999999999999999\textwidth}\kern-0.25em}}}{\colorbox[rgb]{0.8330642160247355,0.937393313996932,0.927089585276211}{\textcolor[rgb]{0,0,0}{\kern-0.25em\hspace*{0.006999999999999999\textwidth}\kern-0.25em}}}{\colorbox[rgb]{0.8392156958580017,0.939607846736908,0.9317647099494935}{\textcolor[rgb]{0,0,0}{\kern-0.25em\hspace*{0.006999999999999999\textwidth}\kern-0.25em}}}{\colorbox[rgb]{0.8453671756912681,0.9418223794768839,0.9364398346227758}{\textcolor[rgb]{0,0,0}{\kern-0.25em\hspace*{0.006999999999999999\textwidth}\kern-0.25em}}}{\colorbox[rgb]{0.8515186555245343,0.9440369122168597,0.9411149592960582}{\textcolor[rgb]{0,0,0}{\kern-0.25em\hspace*{0.006999999999999999\textwidth}\kern-0.25em}}}{\colorbox[rgb]{0.8576701353578007,0.9462514449568356,0.9457900839693406}{\textcolor[rgb]{0,0,0}{\kern-0.25em\hspace*{0.006999999999999999\textwidth}\kern-0.25em}}}{\colorbox[rgb]{0.8638216151910669,0.9484659776968114,0.950465208642623}{\textcolor[rgb]{0,0,0}{\kern-0.25em\hspace*{0.006999999999999999\textwidth}\kern-0.25em}}}{\colorbox[rgb]{0.8730488349409664,0.9517877768067753,0.9574778956525466}{\textcolor[rgb]{0,0,0}{\kern-0.25em\hspace*{0.006999999999999999\textwidth}\kern-0.25em}}}{\colorbox[rgb]{0.8792003147742328,0.9540023095467511,0.9621530203258291}{\textcolor[rgb]{0,0,0}{\kern-0.25em\hspace*{0.006999999999999999\textwidth}\kern-0.25em}}}{\colorbox[rgb]{0.885351794607499,0.956216842286727,0.9668281449991114}{\textcolor[rgb]{0,0,0}{\kern-0.25em\hspace*{0.006999999999999999\textwidth}\kern-0.25em}}}{\colorbox[rgb]{0.8915032744407654,0.9584313750267028,0.9715032696723938}{\textcolor[rgb]{0,0,0}{\kern-0.25em\hspace*{0.006999999999999999\textwidth}\kern-0.25em}}}{\colorbox[rgb]{0.8976547542740317,0.9606459077666788,0.9761783943456762}{\textcolor[rgb]{0,0,0}{\kern-0.25em\hspace*{0.006999999999999999\textwidth}\kern-0.25em}}}{\colorbox[rgb]{0.9021914706510656,0.9623990795191597,0.9773933116127463}{\textcolor[rgb]{0,0,0}{\kern-0.25em\hspace*{0.006999999999999999\textwidth}\kern-0.25em}}}{\colorbox[rgb]{0.9066205361310173,0.9641214938724743,0.9783775483860689}{\textcolor[rgb]{0,0,0}{\kern-0.25em\hspace*{0.006999999999999999\textwidth}\kern-0.25em}}}{\colorbox[rgb]{0.911049601610969,0.9658439082257888,0.9793617851593915}{\textcolor[rgb]{0,0,0}{\kern-0.25em\hspace*{0.006999999999999999\textwidth}\kern-0.25em}}}{\colorbox[rgb]{0.9154786670909209,0.9675663225791034,0.9803460219327141}{\textcolor[rgb]{0,0,0}{\kern-0.25em\hspace*{0.006999999999999999\textwidth}\kern-0.25em}}}{\colorbox[rgb]{0.9199077325708725,0.9692887369324179,0.9813302587060367}{\textcolor[rgb]{0,0,0}{\kern-0.25em\hspace*{0.006999999999999999\textwidth}\kern-0.25em}}}{\colorbox[rgb]{0.9243367980508244,0.9710111512857325,0.9823144954793593}{\textcolor[rgb]{0,0,0}{\kern-0.25em\hspace*{0.006999999999999999\textwidth}\kern-0.25em}}}{\colorbox[rgb]{0.9287658635307761,0.9727335656390471,0.9832987322526819}{\textcolor[rgb]{0,0,0}{\kern-0.25em\hspace*{0.006999999999999999\textwidth}\kern-0.25em}}}{\colorbox[rgb]{0.9331949290107279,0.9744559799923617,0.9842829690260045}{\textcolor[rgb]{0,0,0}{\kern-0.25em\hspace*{0.006999999999999999\textwidth}\kern-0.25em}}}{\colorbox[rgb]{0.9376239944906796,0.9761783943456762,0.9852672057993271}{\textcolor[rgb]{0,0,0}{\kern-0.25em\hspace*{0.006999999999999999\textwidth}\kern-0.25em}}}{\colorbox[rgb]{0.9420530599706313,0.9779008086989908,0.9862514425726497}{\textcolor[rgb]{0,0,0}{\kern-0.25em\hspace*{0.006999999999999999\textwidth}\kern-0.25em}}}{\colorbox[rgb]{0.9464821254505831,0.9796232230523053,0.9872356793459724}{\textcolor[rgb]{0,0,0}{\kern-0.25em\hspace*{0.006999999999999999\textwidth}\kern-0.25em}}}{\colorbox[rgb]{0.9509111909305348,0.9813456374056199,0.988219916119295}{\textcolor[rgb]{0,0,0}{\kern-0.25em\hspace*{0.006999999999999999\textwidth}\kern-0.25em}}}{\colorbox[rgb]{0.9553402564104866,0.9830680517589345,0.9892041528926176}{\textcolor[rgb]{0,0,0}{\kern-0.25em\hspace*{0.006999999999999999\textwidth}\kern-0.25em}}}{\colorbox[rgb]{0.9597693218904383,0.9847904661122491,0.9901883896659402}{\textcolor[rgb]{0,0,0}{\kern-0.25em\hspace*{0.006999999999999999\textwidth}\kern-0.25em}}}{\colorbox[rgb]{0.9641983873703901,0.9865128804655636,0.9911726264392628}{\textcolor[rgb]{0,0,0}{\kern-0.25em\hspace*{0.006999999999999999\textwidth}\kern-0.25em}}}\hspace{0.5em}\raisebox{-0.25em}{$\rightarrow$ \textsf{better}}}
\vspace*{0.5em}
\end{table}

\begin{table}[p]
\centering
\begin{tabular*}{\columnwidth}{l@{\extracolsep{\stretch{1}}}cccccccc@{}}
\toprule
\textbf{Group} & \SAFEnonelzwbyte & \SAFElzwuniformbyte & \SAFElzwuniformstoken & \SAFElzwpolyastokenuniformbyte & \SAFEppmtraininguniformbyte & \SAFEppmtraininguniformstoken & \SAFEppmtrainingpolyastokenuniformbyte & \SAFEppmii\\
\midrule
\ascii & 3.448 & 3.428 & 3.508 & 3.418 & 2.350 & 2.410 & 2.320 & \textbf{2.236}\\
\utfeight & 2.841 & 2.839 & 2.527 & 2.492 & 1.980 & 1.895 & \textbf{1.860} & 1.996\\
Binary & 2.462 & 2.467 & 2.508 & 2.474 & 1.715 & 1.709 & 1.675 & \textbf{1.389}\\
All & 3.010 & 3.002 & 2.881 & 2.826 & 2.078 & 2.057 & 2.002 & \textbf{2.001}\\
\bottomrule
\end{tabular*}
\caption[Mean compression effectiveness]{Mean compression effectiveness, in \bitsperbyte, over the groups and compressors in \cref{table:eval:per-file}. We also include \nonelzwbyte, a \lzw compressor equivalent to \lzwuniformbyte without escaping. All figures are given to 3 decimal places. The best compressor in each row is in \textbf{bold}.}
\label{table:eval:summary}
\end{table}

We test the hypothesis that a token alphabet improves the compression effectiveness of \lzw and \ppm on \utfeight text.
\pagebreak %

We made three variants of each algorithm:
one with a uniform base distribution over the byte alphabet~(\uniformbyte),
one with a uniform base distribution over the token alphabet~(\uniformtoken),
and one with the P\'olya tree base model over the token alphabet~(\polyatoken).
The resulting six algorithms are called
\lzwuniformbyte, \lzwuniformstoken, \lzwpolyastokenuniformbyte,
and \ppmuniformbyte, \ppmuniformtoken, \ppmpolyatoken.

The results in \cref{table:eval:per-file} show that using a P\'{o}lya tree base model improves compression effectiveness: \lzwpolyastokenuniformbyte and \ppmtrainingpolyastokenuniformbyte outperform \lzwuniformstoken and \ppmtraininguniformstoken on every file. We therefore focus on \lzwpolyastokenuniformbyte and \ppmtrainingpolyastokenuniformbyte in the remainder of this section.

We first implemented a traditional version of \lzw, denoted \nonelzwbyte, and then extended it to support escaping. \Cref{table:eval:summary} shows that on average the escaped version, \lzwuniformbyte, is a little better on text files than the original, \nonelzwbyte.
This is probably because \lzwuniformbyte's dictionary can avoid allocating entries for byte values that do not occur.
On binary files, where all byte values may occur, \nonelzwbyte performs slightly better than \lzwuniformbyte.

However, the real benefit of using escaping in \lzw comes from operating over tokens.
On \utfeight texts, \lzwpolyastokenuniformbyte outperforms byte-based \lzwuniformbyte on every file, by an average of 0.347 bits/byte.
On \ascii and binary files, their performance is the same to within 0.010 bits/byte, with \lzwpolyastokenuniformbyte winning on \ascii texts but losing on binary files.

The results are broadly similar for \ppm.
\ppmtrainingpolyastokenuniformbyte outperforms \ppmtraininguniformbyte on all \utfeight texts, by an average of 0.120 bits/byte.
Surprisingly, \ppmtrainingpolyastokenuniformbyte also wins on all but one of the \ascii and binary files, with an average improvement of 0.030 bits/byte and 0.040 bits/byte respectively.

To enable a direct comparison between the byte and token alphabets, our \ppm implementation intentionally omits many refinements to the original \ppm algorithm.
Nonetheless, on \utfeight texts we found it to outperform \ppmii, with \ppmtrainingpolyastokenuniformbyte winning on nine out of twelve files and giving an average improvement of 0.136 bits/byte.
However, \ppmii wins on all \ascii texts and binary files.
Interestingly, \ppmtraininguniformbyte often outperforms \ppmii on \utfeight texts (although by a slimmer margin than \ppmtrainingpolyastokenuniformbyte), which suggests that \ppmii may have been tuned for texts encoded in 8-bit alphabets, rather than multi-byte encodings such as \utfeight.

\Section{Conclusions}

We extended the byte-based compressors \lzw and \ppm to operate over \emph{tokens}: combinations of Unicode characters and byte values for error conditions.
Our variants substantially outperform the original algorithms on a \utfeight corpus that we hope is somewhat representative of the world's current use of languages.
Furthermore, \ppmtrainingpolyastokenuniformbyte (our version of \ppm) is more effective than \ppmii (an extensively tuned variant of \ppm) on most texts in this corpus. In contrast to the existing Unicode compressors \scsu and \bocu, our compressors can handle all inputs, including \utfeight text embedded inside binary formats.

Despite the radically different designs of \lzw and \ppm, both compressors enjoy a similar improvement in compression effectiveness from operating over tokens. This result suggests that other compressors might also benefit from our technique.

Our source code and data sets are open source and available to download from \url{https://github.com/AdamGleave/UnicodeCompressor}.

\Section*{Acknowledgements}
The authors wish to thank Zoubin Ghahramani for helpful discussions and feedback.

\Section*{References}
\bibliographystyle{IEEEtran}
\bibliography{refs}

\end{document}